\documentstyle[12pt]{article}
\newcommand{\be}{\begin{equation}}
\newcommand{\bea}{\begin{eqnarray}}
\newcommand{\ee}{\end{equation}}
\newcommand{\eea}{\end{eqnarray}}

\begin{document}
\topmargin -1cm
\oddsidemargin=0.25cm\evensidemargin=0.25cm
\renewcommand{\thefootnote}{\fnsymbol{footnote}}
\thispagestyle{empty}
\begin{center}
{\large\bf On the Schr\"odinger Equation for the Minisuperspace Models }
\vspace{0.5cm} \\
{\bf V.I. Tkach}\footnote{E-mail: vladimir@ifug1.ugto.mx}
\vspace{0.5cm} \\
{\it Instituto de F\'{\i}sica, Universidad de Guanajuato, \\
Apartado Postal E-143, C.P. 37150, Le\'on, Gto. M\'exico}
\vspace{0.5cm}\\
\end{center}
\begin{center}
{\bf A. Pashnev}\footnote{E-mail: pashnev@thsun1.jinr.ru} and {\bf
J.J. Rosales}\footnote{E-mail: rosales@thsun1.jinr.ru}
\vspace{0.5cm}\\ {\it JINR-Bogoliubov Laboratory of Theoretical
Physics,\\ 141980 Dubna, Moscow Region, Russia} \vspace{1.5cm}\\
\end{center}
\begin{abstract}
We obtain a time-dependent Schr\"odinger equation for the
Friedmann - Robertson - Walker (FRW) model interacting with a
homogeneous scalar matter field. We show that for this purpose it
is necesary to include an additional action invariant under the
reparametrization of time. The last one does not change the
equations of motion of the system, but changes only the constraint
which at the quantum level becomes time-dependent Schr\"odinger
equation. The same procedure is applied to the supersymmetric case
and the supersymmetric quantum constraints are obtained, one of
them is a square root of the Schr\"odinger operator.

\end{abstract}

\newpage
\setcounter{page}1

One of the most important questions in quantum cosmology is that of
identifying a suitable time parameter \cite{1} and a time-dependent
Wheeler-DeWitt equation \cite{2,3}. The main peculiarity of the
gravity theory is the presence of non-physical variables (gauge variables)
and constraints \cite{3,4,5,6}. They arise due to the general coordinate
invariance of the theory. The conventional Wheeler-DeWitt formulation
gives a time independent quantum theory \cite{7}. The canonical quantization
of the minisuperspace approximation \cite{8} has been used to find results
in the hope, that they would illustrate the behaviour of general relativity
\cite{9}. In the minisuperspace models \cite{2,7} there is a residual
invariance under reparametrization of time (world-line symmetry). Due to
this fact the equation that governs the quantum behaviour of these models
is the Schr\"odinger equation for states with zero energy. On the other hand,
supersymmetry transformations are more fundamental than time translations
(reparametrization of time) in the sense, that these ones may be generated
by anticommutators of the supersymmetry generators. The recent introduction
of supersymmetric mini\-superspace models has led to the square root
equations for states with zero energy \cite{10,11,12}. The
structure of the world-line supersymmetry or the world-line supersymmetry
transformations has led to the zero Hamiltonian phenomena \cite{2,6,12}.
Investigations of the time evolution problem for such quantum systems
have been carried in two directions: the cosmological models of gravity
have been quantized by reducing the phase space degrees of freedom
\cite{13,14,15,16,17} and with the help of the WKB approach \cite{18,19}.

In this work we obtain a time-dependent Schr\"odinger equation for the
homogeneous cosmological models. In our approach this equation arises due
to an additional action invariant under reparametrization. The last one
does not change the equations of motion, but the constraint which becomes
time-dependent Schr\"odinger equation. In the case of the supersymmetric
minisuperspace model we obtain the supersymmetric constraints, one of them
is a square root of time-dependent Schr\"odinger equation.

We begin by considering an homogeneous and isotropic metric defined by
\be
d s^2 = - N^2 (t) dt^2 +R^2 (t) d\Omega^2_3 ,
\label{1}
\ee
where the only dynamical degree of freedom is the scale factor $R(t)$.
The lapse function $N(t)$, being a pure gauge variable, is not dynamical.
The quantity $d \Omega^2_3$ is the standard line element on the unit
three-sphere. We shall set $c= \hbar=1$. The pure gravitational action
corresponding to the metric (\ref{1}) is
\be
S_g= \frac{6}{\kappa^2} \int \left( -\frac{R\dot R^2}{2N} +\frac{1}{2}
k NR \right) dt ,
\label{2}
\ee
where $k =1,0,-1$ corresponds to a closed, flat or open space.
$\kappa^2 = 8\pi G_N $, where $G_N $ is the Newton's constant of
gravity, and the overdot denotes differentiation with respect to $t$. The
action (\ref{2}) preserves the invariance under the time reparametrization
\be
t^\prime \to t + a (t),
\label{3}
\ee
if the transformations of $N(t)$ and $R(t)$ are
\be
\delta R = a\dot R \qquad\qquad \delta N= \dot a N + a\dot N
\label{4}
\ee
that is, $R(t)$ transforms as a scalar and $N(t)$ as a one-dimensional
vector, and its dimensionality is the inverse of $a(t)$.

So, we consider the interacting action for the homogeneous real scalar
matter field $\phi(t)$ and the scale factor $R(t)$. This action has the form
\be
S_m = \int \left( \frac{R^3 \dot\phi^2}{2N} - NR^3 V(\phi) \right) dt.
\label{5}
\ee
This action remains invariant under the local transformation (\ref{3}), if
in addition to the transformation law for $R(t)$ and $N(t)$ in (\ref{4}), the
field $\phi(t)$ transforms as a scalar; $\delta\phi = a \dot\phi$.

Thus, our system is described by the full action
\be
S= S_g + S_m = \int \left( -\frac{3R{\dot R}^2}{\kappa^2 N} +
\frac{R^3 {\dot \phi}^2}{2N} + \frac{3 k N R}{\kappa^2} -
NR^3 V(\phi) \right)dt.
\label{6}
\ee
Now, we shall consider the Hamiltonian analysis of this action. The
canonical momenta for the variables $R$ and $\phi$ are given,
respectively, by
\be
P_R = \frac{\partial L}{\partial\dot R} = - \frac{6R \dot R}{\kappa^2 N},
\qquad P_{\phi} = \frac{R^3 \dot \phi}{N}.
\label{7}
\ee
Their canonical Poisson brackets are defined as
\be
\lbrace R, P_R \rbrace = 1, \qquad\qquad
\lbrace \phi, P_{\phi} \rbrace = 1.
\label{8}
\ee
The canonical momentum for the variable $N(t)$ is
\be
P_N \equiv \frac{\partial L}{\partial \dot N} = 0,
\label{9}
\ee
this equation merely constrains the variable $N(t)$ (primary constraint).
The canonical Hamiltonian can be calculated in the usual way, it has
the form $H_c = NH_0$, then the total Hamiltonian is
\be
H_T = NH_0 + u_N P_N,
\label{10}
\ee
where $u_N$ is the Lagrange multiplier associated to the constraint
$P_N = 0$ in (\ref{9}), and  $H_0$ is the Hamiltonian written as
\be
H_0 = \left( - \frac{\kappa^2 P^2_R}{12R} + \frac{\pi^2_\phi}{2R^3} -
\frac{3kR}{\kappa^2} + R^3 V(\phi ) \right).
\label{11}
\ee
The time evolution of any dynamical variables is generated by
(\ref{10}). For the compatibility of the constraint the Eq. (\ref{9})
and the dynamics generated by the total Hamiltonian of Eq. (\ref{10}), the
following equation must hold
\be
H_0=0,
\label{12}
\ee
which constrains the dynamics of our system. So, we proceed to the quantum
mechanics from the above classical system. We introduce the wave function
of the Universe $\psi$. The constraint equation (\ref{12}) must be
imposed on the states
\be
 H_0 \psi = 0.
\label{13}
\ee
This constraint nullifies all the dynamical evolution generated by the total
Hamiltonian (\ref{10}). A commutator of any operator and the total
Hamiltonian becomes zero, if it is evaluated for the above constrained states.
The disappearence of time seems disappointing, however, it is a proper
consequence of the invariance of general coordinate transformation in general
relativity. The equation (\ref{9}) merely says, that the wave function $\psi$
does not depend on the lapse function $N(t)$. Therefore, we expect that the
equation in (\ref{13}) may contain any information of dynamics. In quantum
cosmology the constraint (\ref{13}) is well-known as the Wheeler-DeWitt
equation (time-independent Schr\"odinger equation).

In order to get a time-dependent Schr\"odinger equation we shall
regard the following invariant action
\be
S_r = - \frac{1}{\kappa^3} \int R^3 P_T \left(- \frac{dT(t)}{dt} +
N(t)\right) dt,
\label{14}
\ee
where $(T, P_T)$ is a pair of dynamical variables, $P_T$ is the momentum
conjugate to $T$. This action is invariant under reparametrization
(\ref{3}), if $P_T$ and $T$ transform as
\be
\delta P_T = a \dot P_T \qquad\qquad \delta T = a\dot T,
\label{15}
\ee
and $N, R$ as in (\ref{4}).

So, adding the action (\ref{14}) to the action (\ref{6}) we have the total
invariant action $\tilde S = S_g + S_m + S_r$. In the first order form we
get
\be
\tilde S = \int \left\{\dot R P_R + \dot \phi P_{\phi} - NH_0
+ \frac{R^3 P_T}{\kappa^3} \left(\frac{dT(t)}{dt} - N(t)\right) \right \} dt.
\label{16}
\ee
We shall proceed with the canonical quantization of the action (\ref{16}).
We define the canonical momenta $\pi_T$ and $\pi_{P_T}$ corresponding to the
variables $T$ and $P_T$, respectively. We get
\be
\pi_T \equiv \frac{\partial \tilde L}{\partial \dot T} =
 \frac{R^3}{\kappa^3} P_T, \qquad\qquad
\pi_{P_T} \equiv \frac{\partial \tilde L}{\partial \dot P_T} = 0,
\label{17}
\ee
leading to the constraints
\be
\Pi_1 \equiv \pi_T - \frac{R^3}{\kappa^3} P_T = 0, \qquad\qquad
\Pi_2 \equiv \pi_{P_T} = 0.
\label{18}
\ee

So, we define the matrix $C_{AB}$, $(A, B =1,2)$ as a Poisson brackets
between the constraints $C_{AB} = \lbrace \Pi_A, \Pi_B \rbrace$. Then,
we have the following non-zero matrix elements
\be
\lbrace \Pi_{1}, \Pi_{2} \rbrace = -\frac{R^3}{\kappa^3},
\label{19}
\ee
with their inverse matrix elements $(C^{-1})^{1,2} = \frac{\kappa^3}{R^3}$.
The Dirac's brackets $\lbrace , \rbrace^\ast$ are defined by
\be
\lbrace f, g \rbrace^{\ast} = \lbrace f, g \rbrace -
\lbrace f, \Pi_A \rbrace C^{AB} \lbrace \Pi_B, g \rbrace.
\label{20}
\ee
The result of this procedure leads to the non-zero Dirac's bracket relation
\be
\lbrace T, P_T \rbrace^{\ast} =  \frac{\kappa^3}{R^3}.
\label{21}
\ee
Then, the canonical Hamiltonian is
\be
\tilde H_c = N \left(\frac{R^3}{\kappa^3} P_T + H_0 \right ),
\label{22}
\ee
where the Hamiltonian constraint corresponding to the action
(\ref{16}) is
\be
\tilde H =  \frac{ R^3}{\kappa^3} P_T + H_0.
\label{23}
\ee
At the quantum level the Dirac's brackets become commutators
\be
[ T, P_T ] = i\lbrace T, P_T \rbrace^{\ast} = i \frac{\kappa^3}{R^3}.
\label{24}
\ee
So, taking the momentum $P_T$ corresponding to $T$ as
\be
P_T = -i \frac{\kappa^3}{R^3} \frac{\partial}{\partial T},
\label{25}
\ee
the quantum constraint (\ref{23}) becomes quantum equation on the wave
function $\psi$
\be
i\frac{\partial}{\partial T} \psi (T,R,\phi )= H \left(- i\frac{\partial}
{\partial R}, - i\frac{\partial}{\partial\phi},R,\phi \right)\psi.
\label{26}
\ee
Explicitly, we have
\be
i\frac{\partial\psi}{\partial T} = \left[ \frac{\kappa^2}{12} \frac{1}{R^2}
\frac{\partial}{\partial R} \left(R \frac{\partial}{\partial R} \right)
-\frac{3kR}{\kappa^2} -\frac{1}{2R^3} \frac{\partial^2}{\partial\phi^2} +
R^3 V(\phi )\right] \psi.
\label{27}
\ee
This equation is the time-dependent Schr\"odinger equation for
minisuperspace.

Equations of motion are obtained by demanding that the action
$ \tilde S = S_g + S_m + S_r$ is extremal, $i.e.$ the functional derivatives
of $ \tilde S $ must be zero
\be
\frac{\delta \tilde S}{\delta R} = \frac{\delta S_g}{\delta R} +
\frac{\delta S_m}{\delta R} + \frac{\delta S_r}{\delta R} = 0.
\label{28}
\ee
As a consequence of the equation of motion
\be
\frac{\delta \tilde S}{\delta P_T} = \frac{\delta S_r}{\delta P_T} =
\frac{R^3}{\kappa^3} (\dot T-N) = 0,
\label{29}
\ee
the last term in (\ref{28}), $ \frac{3R^2}{\kappa^3} P_T (\dot T - N) $
dissapears and, in fact, inclusion in $ S $ of
an additional invariant action $S_r$ does not change the equations of motion
axcept the equation $\frac{\delta \tilde S}{\delta N} = 0 $,
which is the constraint (\ref{23}).

In the case of the Arnowit-Deser-Misner (ADM) formalism \cite{3} the
additional term (\ref{14}) can be written in the following invariant form
\begin{eqnarray}
S_{(d=4)} &=& - \frac{1}{\kappa^3}\int \sqrt{-g} P_T (-n^\mu \partial_\mu
 T + 1) d^4 x \nonumber\\
&=& - \frac{1}{\kappa^3} \int N h^{1/2} P_T
\left(- \frac{\partial_0 T}{N} - \frac{N^i\partial_i T}{N}
+ 1 \right) dt d^3 x \label{30}\\
&=& - \frac{1}{\kappa^3} \int h^{1/2} P_T (-\partial_0 T -
N^i \partial_i T + N)dt d^3 x. \nonumber
\end{eqnarray}
According to the ADM prescription \cite{3} of classical general relativity,
one considers a slicing of the space-time by a family of space-like
hypersurfaces labeled by a parameter $t$. This parameter can be thought of
as a time coordinate, so that each slice is identified by the relation
$t=const$. The remaining three spatial coordinates, $x^i$, determine a
coordinatization of each slice. The space-time metric $g_{\mu\nu}$ is
decomposed into shift $N^i$, lapse $N$ functions and the three-metric
of the slice $h_{ij}$. In the action (\ref{30}), $h= det h_{ij} $,
$\sqrt {-g} = N\sqrt h$ and $n^\mu$ $(n^\mu n_\mu =-1)$ is the unit normal
vector to hypersurface $t=const$ with components $n_\mu = (-N, 0,0,0)$ and
$n^\mu = \Big(\frac{1}{N}, - \frac{N^i}{N} \Big)$. In the case of the
homogeneous metric (\ref{1}) the shift vector is $N^i =0$ and
$h^{1/2} = R^3$.

So, if we consider the four-dimensional gravity interacting with a scalar
matter field and the invariant additional term (\ref{30}), then after
applying the (ADM) $(3 + 1)$ formalism for the FRW model we get
\begin{eqnarray}
S &=& - \frac{1}{2\kappa^2} \int \sqrt{-g} R d^4 x - \int \sqrt{-g}
\left[\frac{(\partial_\mu \phi )^2}{2} + V(\phi ) \right] d^4 x
\nonumber\\
&-& \frac{1}{\kappa^3} \int \sqrt{-g} P_T (n^\mu \partial_\mu T-1) d^4 x
= \int \left[ \left (- \frac{3R\dot R^2}{2N\kappa^2} + \frac{3}{\kappa^2}
kNR \right) + \right.\label{32}\\
&+& \left. \frac{R^3 \dot\phi^2}{2N} - NR^3V(\phi ))\right] dt
+ \frac{1}{\kappa^3} \int P_T R^3 \left( \frac{dT(t)}{dt} -
N(t) \right) dt. \nonumber
\end{eqnarray}
In particular, choosing the gauge $N=1$, then $T=t$ and we obtain the
so-called cosmic time, on the other hand, if we take
$N= \frac{R}{\kappa}$ we get the conformal time gauge.

In order to obtain a superfield formulation of the action (\ref{6})
the transformation of the time reparametrization (\ref{3}) must
be extended to the $n=2$ local conformal time supersymmetry (LCTS)
$(t, \eta, \bar\eta)$ \cite{20,21}. These (LCTS) transformations can be
written as
\begin{eqnarray}
\delta t &=& {I\!\!L}(t, \theta,\bar\theta) + \frac{1}{2}\bar\theta
D_{\bar\theta} {I\!\!L}(t,\theta,\bar\theta) - \frac{1}{2}\theta D_{\theta}
{I\!\!L}(t,\theta,\bar\theta), \nonumber\\
\delta \theta &=& \frac{i}{2} D_{\bar\theta} {I\!\!L}(t,\theta,\bar\theta),
\qquad \delta \bar\theta = -\frac{i}{2} D_{\theta}
{I\!\!L}(t,\theta,\bar\theta),
\label{33}
\end{eqnarray}
with the superfunction ${I\!\!L}(t,\theta,\bar\theta)$ defined by
\be
{I\!\!L}(t,\theta,\bar\theta) = a(t) + i\theta {\bar\beta}^\prime(t) +
i\bar\theta \beta^\prime(t) + b(t)\theta \bar\theta,
\label{34}
\ee
where $D_{\theta} = \frac{\partial}{\partial \theta} + i\bar\theta
\frac{\partial}{\partial t}$ and $D_{\bar\theta} = - \frac{\partial}
{\partial \bar\theta} - i\theta \frac{\partial}{\partial t}$ are the
supercovariant derivatives of the $n=2$ supersymmetry, $a(t)$ is a local
time reparametrization parameter, $\beta^\prime(t)= N^{-1/2} \beta$ is the
Grassmann complex parameter of the local conformal $n=2$ supersymmetry
transformations and $b(t)$ is the parameter of the local $U(1)$ rotations
on the Grassmann coordinates $\theta$ $(\bar\theta = \theta^\dagger)$.
Then, the superfield generalization of the action (\ref{6}), which is
invariant under the $n=2$ (LCTS) transformations (\ref{33}) has the form
\cite{22,23}
\begin{eqnarray}
S_{(n=2)} &=& S_g + S_m = \int \left( -\frac{3}{\kappa^2}{I\!\!N}^{-1}
{I\!\!R} D_{\bar\theta} {I\!\!R} D_{\theta}{I\!\!R} +
\frac{3 \sqrt k}{\kappa^2} {I\!\!R}^2 \right)
d\theta d\bar\theta dt \label{35}\\
&+& \int \left (\frac{1}{2}{I\!\!N}^{-1} {I\!\!R}^3 D_{\bar\theta}
{\bf \Phi} D_{\theta} {\bf \Phi} - 2 {I\!\!R}^3 g(\bf \Phi) \right )
d\theta d\bar\theta dt, \nonumber
\end{eqnarray}
where $g(\Phi)$ is the superpotential. The most general supersymmetric
interaction for the set of complex homogeneous scalar fields with the
scale factor was considered in \cite{24,25}. For the one-dimensional
gravity superfield ${I\!\!N}(t,\theta,\bar\theta)$ we have the following
series expansion
\be
{I\!\!N}(t,\theta,\bar\theta) = N(t) + i\theta {\bar\psi}^\prime(t) +
i\bar\theta \psi^\prime(t) + V^\prime(t) \theta\bar\theta,
\label{36}
\ee
where $N(t)$ is the lapse function, $\psi^\prime(t) = N^{1/2}(t) \psi(t)$
and $V^\prime(t) = N(t)V(t) + \bar\psi(t) \psi(t)$. The components
$N, \psi, \bar\psi$ and $V$ in (\ref{36}) are gauge fields of the
one-dimensional $n=2$ supergravity. The superfield (\ref{36}) transforms as
the one-dimensional vector under the (LCTS) transformations (\ref{33}),
\be
\delta {I\!\!N} = ({I\!\!L}{I\!\!N})^. + \frac{i}{2}D_{\bar\theta}
{I\!\!L}D_{\theta} {I\!\!N} + \frac{i}{2}D_{\theta}{I\!\!L} D_{\bar\theta}
{I\!\!N}.
\label{37}
\ee
The series expansion for the superfield ${I\!\!R}(t,\theta,\bar\theta)$ has
a similar form
\be
{I\!\!R}(t,\theta,\bar\theta) = R(t) + i\theta {\bar\lambda}^\prime(t) +
i\bar\theta \lambda^\prime(t) + B^\prime(t) \theta\bar\theta,
\label{38}
\ee
where $R(t)$ is the scale factor of the FRW Universe,
$\lambda^\prime = \kappa N^{1/2} \lambda$ and
$B^\prime(t)= \kappa N(t) B(t) + \frac{\kappa}{2}(\bar\psi(t) \lambda(t) -
\psi(t) \bar\lambda(t))$.

For the real scalar matter superfield ${\bf \Phi}(t,\theta,\bar\theta)$ we
have
\be
{\bf \Phi}(t,\theta,\bar\theta) = \phi(t) + i\theta \bar\chi^\prime(t) +
i\bar\theta \chi^\prime(t) + F^\prime(t) \theta\bar\theta,
\label{39}
\ee
where $\chi^\prime(t) = N^{1/2}(t)\chi(t)$ and
$F^\prime(t) = N(t)F(t) + \frac{1}{2}(\bar\psi(t) \bar\chi(t) -
\psi(t) \chi(t))$. The components $B(t)$ and $F(t)$ in the superfields
${I\!\!R}$ and ${\bf \Phi}$ are auxiliary fields. The superfields (\ref{38})
and (\ref{39}) transform as scalars under the (LCTS) transformations
(\ref{33}).

Performing the integration over $\theta, \bar\theta$ in (\ref{35}) and
eliminating the auxiliary fields $B$ and $F$ by means of their equations of
motion, the action (\ref{35}) takes its component form. The first-class
constraints may be obtained from the component form of the action
(\ref{35}) varying it with respect to $N(t), \psi(t), \bar\psi(t)$ and
$V(t)$, respectively. Then, we obtain the following first-class constraints
$H_0=0$, $S=0$, $\bar S= 0$ and $F=0$, where
\begin{eqnarray}
H_0 &=& - \frac{\kappa^2}{12} \frac{\pi^2_R}{R} - \frac{3kR}{\kappa^2}
-\frac{\sqrt k}{3R} \bar\lambda \lambda + \frac{\pi^2_\phi}{2R^3} -
\frac{i\kappa}{2R^3} \pi_\phi (\bar\lambda \chi + \lambda \bar\chi)
- \frac{\kappa^2}{4R^3} \bar\lambda \lambda \bar\chi \chi \nonumber\\
&+& \frac{3 \sqrt k}{2R} \bar\chi \chi + \kappa^2 g(\phi) \bar\lambda \lambda
+ 6 \sqrt k g(\phi)R^2 + 2 \left(\frac{\partial g}{\partial \phi}\right)^2
R^3 - 3 \kappa^2 g^2(\phi) R^3 \nonumber\\
&+& \frac{3}{2} \kappa^2 g(\phi) \bar\chi\chi
+ 2 \frac{\partial^2 g}{\partial \phi^2} \bar\chi \chi +
\kappa \frac{\partial g}{\partial \phi}(\bar\lambda \chi - \lambda \bar\chi),
\label{40}
\end{eqnarray}
\begin{eqnarray}
S&=& \left(\frac{i\kappa}{3} R^{-1/2} \pi_R - \frac{2\sqrt k }
{\kappa}R^{1/2} + 2\kappa g(\phi) R^{3/2} + \frac{\kappa}{4} R^{-3/2}
\bar\chi \chi \right) \lambda \nonumber\\
&+& \left( iR^{-3/2} \pi_\phi + 2 R^{3/2}
\frac{\partial g}{\partial \phi}\right ) \chi, \label{41}\\
\bar S &=& S^{\dagger}, \nonumber
\end{eqnarray}
and
\be
F = - \frac{2}{3}\bar\lambda \lambda + \bar\chi \chi,
\label{42}
\ee
The canonical Hamiltonian is the sum of all the
constraints
\be
H_{c(n=2)} = NH_0 + \frac{1}{2} \bar\psi S - \frac{1}{2}\psi \bar S +
\frac{1}{2} V F.
\label{43}
\ee
In terms of Dirac's brackets for the canonical variables $R, \pi_R, \phi,
\pi_\phi, \lambda, \bar\lambda, \chi$ and $\bar\chi$ the quantities $H_0,
S,\bar S$ and $F$ form the closed super-algebra of conserving charges
\begin{eqnarray}
\lbrace S, \bar S \rbrace^\ast &=& - 2iH_0, \qquad
\lbrace H_0, S \rbrace^\ast = \lbrace H_0, \bar S \rbrace^\ast = 0
\label{44} \\
\lbrace F, S\rbrace^\ast &=& iS, \qquad
\lbrace F, \bar S \rbrace^\ast = -i\bar S. \nonumber
\end{eqnarray}
So, any physically allowed states must obey the following quantum constraints
\begin{eqnarray}
H_0 \psi &=& 0, \qquad S\psi = 0, \qquad \bar S \psi = 0, \qquad
F \psi = 0,
\label{45}
\end{eqnarray}
when we change the classical variables by their corresponding
operators. The first equation in (\ref{45}) is the Wheeler-DeWitt
equation for the minisuperspace model. Therefore, we have the {\it
time-independent} Schr\"odinger equation, this fact is due to the
invariance of the action (\ref{35}) under reparametrization
symmetry, this problem is well-known as the ``problem of time"
\cite{1} in the minisuperspace models and general relativity
theory. Due to the super-algebra (\ref{44}) the second and the
thirth equations in (\ref{45}) reflect the fact, that there is a
``square root" of the Hamiltonian $H_0$ with zero energy states.
The constraints Hamiltonian $H_0$, supercharges $S, \bar S$ and
the fermion number operator $F$ follow from the invariance of the
action (\ref{35}) under the $n=2$ (LCTS) transformations
(\ref{33}).

In order to have a time-dependent Schr\"odinger equation for the
supersymmetric minisuperspace models with the action (\ref{35})
we consider a generalization of the reparametrization invariant action
$S_r$ (\ref{14}). In the case of $n=2$ (LCTS) it has the superfield form
\begin{eqnarray}
S_{r(n=2)} &=& \int \left[ {I\!\!P} -\frac{i}{2} {I\!\!N}^{-1}
\left( D_{\bar\theta} {\bf T} D_\theta{I\!\!P} - D_{\bar\theta}
{I\!\!P}D_\theta {\bf T} \right)\right] d\theta d\bar\theta dt.
\label{46}
\end{eqnarray}
Note, that the $Ber E^A_B$, as well as the superjacobian of $n=2$
(LCTS) transformations, is equal to one and is omitted in the actions
(\ref{35},\ref{46}). The action (\ref{46}) is determined in terms of the new
superfields ${\bf T}$ and ${I\!\!P}$. The series expansion for ${\bf T}$
has the form
\be
{\bf T}(t,\theta,\bar\theta) = T(t) + \theta \eta^\prime(t) -
\bar\theta \bar\eta^\prime(t) + m^\prime(t) \theta\bar\theta,
\label{47}
\ee
where $ \eta^\prime(t) = N^{1/2}(t) \eta(t)$ and $m^\prime(t) =
N(t)m(t) + \frac{i}{2}(\bar\psi(t) \bar\eta(t) + \psi(t) \eta(t))$.
The superfield ${\bf T}$ is determined by the odd complex time variables
$\eta(t)$ and $\bar\eta(t)$, which are the superpartners of the time
$T(t)$ and one auxiliary parameter $m(t)$. The transformation rule for the
superfield ${\bf T}(t,\theta,\bar\theta)$ under the $n=2$ (LCTS)
transformations (\ref{33}) is
\be
\delta {\bf T} = {I\!\!L} \dot{\bf T} + \frac{i}{2}D_{\bar\theta}
{I\!\!L} D_\theta {\bf T} + \frac{i}{2}D_\theta {I\!\!L}D_{\bar\theta}
{\bf T}.
\label{48}
\ee
The superfield ${I\!\!P}(t,\theta,\bar\theta)$ has the form
\be
{I\!\!P}(t,\theta,\bar\theta) = \rho(t) + i\theta P^\prime_{\bar\eta}(t)
+ i\bar\theta P^\prime_\eta(t) + P^\prime_T(t) \theta\bar\theta,
\label{49}
\ee
where $P^\prime_\eta (t) = N^{1/2} P_\eta$ and $P^\prime_T (t) =
N P_T + \frac{1}{2}(\bar\psi P_{\eta} - \psi P_{\bar\eta})$,
$P_\eta $ and $P_{\bar\eta}$ are the odd complex momenta, $i.e.$ the
superpartners of the momentum $P_T$.

The superfield ${I\!\!P}(t,\theta,\bar\theta)$ transforms as
\be
\delta {I\!\!P}(t,\theta,\bar\theta) = {I\!\!L} \dot {I\!\!P} +
\frac{i}{2}D_{\bar\theta} {I\!\!L} D_\theta {I\!\!P} +
\frac{i}{2} D_\theta {I\!\!L} D_{\bar\theta} {I\!\!P}.
\label{50}
\ee
The action (\ref{46}) is invariant under the $n=2$ (LCTS) transformations
(\ref{33}). Performing the integration over $\theta$ and $\bar\theta$ in
(\ref{46}) and making the redefinitions $P_T \to \frac{R^3}{\kappa^3}P_T$,
$P_\eta \to \frac{R^3}{\kappa^3}P_\eta$ and
$P_{\bar\eta} \to \frac{R^3}{\kappa^3}P_{\bar\eta}$ we obtain its component
form
\begin{eqnarray}
S_{r(n=2)} &=& -\int \left\{ \frac{R^3}{\kappa^3} \left(P_T(N - \dot T) +
i\dot \eta P_\eta + i\dot{\bar\eta} P_{\bar\eta} +
\frac{\bar\psi}{2}(P_\eta - \bar\eta P_T) \right. \right.\label{51}\\
&-& \left. \left. \frac{\psi}{2}(P_{\bar\eta} - \eta P_T) +
\frac{V}{2}(\eta P_\eta - \bar\eta P_{\bar\eta})\right) + m\dot \rho
-\frac{iR^3}{2\kappa^3}m\psi P_{\bar\eta} - \frac{iR^3}{2\kappa^3}
m \bar\psi P_\eta \right\} dt.
\nonumber
\end{eqnarray}
We can see from (\ref{51}) that the momenta $P_{\eta}$, $P_{\bar\eta}$ and
$P_T$ in the superfield (\ref{49}) are related with the components of the
superfield (\ref{36}), which enter in the action (\ref{35}), unlike those
momenta, the component $\rho$ of the superfield (\ref{49}) is not related
with any components in (\ref{36}). Therefore, the variables $\rho$ and
$m$ can be eliminated from the action (\ref{51}) by means of their equations
of motion, then the component action has the final form
\begin{eqnarray}
S_{r(n=2)} &=& -\int \frac{R^3}{\kappa^3}\left\{ P_T(N - \dot T) +
i\dot \eta P_\eta + i\dot{\bar \eta}P_{\bar\eta} + \frac{\bar\psi}{2}
(P_\eta - \bar\eta P_T) \right. \nonumber\\
&-& \left. \frac{\psi}{2}(P_{\bar\eta} - \eta P_T) +
\frac{V}{2}(\eta P_\eta - \bar\eta P_{\bar\eta}) \right\} dt.
\label{52}
\end{eqnarray}
In addition to the canonical momenta $\pi_T$ and $\pi_{P_T}$
for the two even constraints (\ref{17}), the action (\ref{52}) has
the additional momenta ${\cal P}_\eta$ and ${\cal P}_{P_\eta}$
conjugate to $\eta$ and $P_\eta$, respectively,
\begin{eqnarray}
{\cal P}_\eta &=& \frac{\partial L_{r(n=2)}}{\partial\dot\eta} =
-i \frac{R^3}{\kappa^3} P_\eta, \qquad
{\cal P}_{P_\eta} = \frac{\partial L_{r(n=2)}}{\partial \dot P_\eta} = 0.
\label{53}
\end{eqnarray}
With respect to the canonical odd Poisson brackets we have
\be
\lbrace \eta , {\cal P}_\eta \rbrace = 1, \qquad
\lbrace P_\eta , {\cal P}_{P_\eta} \rbrace = 1.
\label{54}
\ee
They form two primary constraints of second-class
\be
\Pi_3 (\eta ) \equiv {\cal P}_\eta +i\frac{R^3}{\kappa^3} P_\eta = 0 ,
\qquad \Pi_4 (P_{\eta})\equiv {\cal P}_{P_{\eta}} = 0.
\label{55}
\ee
The only non-vanishing Poisson bracket between these constraints is
\be
\lbrace \Pi_3 , \Pi_4 \rbrace = i \frac{R^3}{\kappa^3} .
\label{56}
\ee
The momenta ${\cal P}_{\bar\eta}$ and ${\cal P}_{P_{\bar\eta}}$ conjugate to
$\bar\eta$ and $P_{\bar\eta}$ respectively, also give two primary
constraints of second-class
\be
\Pi_5 (\bar\eta) \equiv {\cal P}_{\bar\eta} + i\frac{R^3}{\kappa^3}
P_{\bar\eta} = 0 ,\qquad \Pi_6 (P_{\bar\eta}) =
{\cal P}_{P_{\bar\eta}} = 0,
\label{57}
\ee
with non-vanishing Poisson bracket
\be
\lbrace \Pi_5, \Pi_6 \rbrace = i \frac{R^3}{\kappa^3}.
\label{58}
\ee
The constraints (\ref{55}) and (\ref{57}) for the Grassmann dynamical
variables can be eliminated by Dirac's procedure. Defining the matrix
constraint $C_{ik} (i,k =\eta ,P_\eta ,\bar\eta ,P_{\bar\eta})$ as the
odd Poisson bracket we have the following non-zero matrix elements
\begin{eqnarray}
C_{\eta P_\eta} &=& C_{P_\eta \eta} = \lbrace \Pi_3,
\Pi_4 \rbrace = i\frac{R^3}{\kappa^3}, \nonumber\\
C_{{\bar\eta} P_{\bar\eta}} &=& C_{P_{\bar\eta\bar\eta}}
= \lbrace \Pi_5, \Pi_6 \rbrace = i \frac{R^3}{\kappa^3},
\label{59}
\end{eqnarray}
with their inverse matrices $(C^{-1})^{\eta P_\eta}= -i \frac{\kappa^3}{R^3}$
and $(C^{-1})^{\bar\eta P_\eta}= -i \frac{\kappa^3}{R^3}$.
The result of this procedure is the elimination of the momenta conjugate to
the Grassmann variables, leaving us with the following non-zero Dirac's
bracket relations
\be
\lbrace \eta ,P_\eta \rbrace^\ast = i\frac{\kappa^3}{R^3}, \qquad
\lbrace \bar\eta , P_{\bar\eta}\rbrace^\ast =  i\frac{\kappa^3}{R^3}.
\label{60}
\ee
So, if we take the additional term (\ref{46}), then the full action is
\be
\tilde S_{(n=2)} = S_{(n=2)} + S_{r(n=2)}.
\label{61}
\ee
The canonical Hamiltonian for the action (\ref{61}) will have the
following form
\begin{eqnarray}
\tilde H_{c(n=2)} &=& N\left (\frac{R^3}{\kappa^3} P_T + H_0 \right)
+ \frac{\bar\psi}{2} \left (\frac{R^3}{\kappa^3} S_\eta + S \right)\nonumber\\
&-& \frac{\psi}{2} \left (-\frac{R^3}{\kappa^3} S_{\bar\eta} + \bar S \right)
+ \frac{V}{2} \left ( \frac{R^3}{\kappa^3} F_\eta + F \right),
\label{62}
\end{eqnarray}
where $S_\eta = (P_\eta - \bar\eta P_T)$, $S_{\bar\eta} =
(- P_{\bar\eta} + \eta P_T)$, $F_\eta =
(\eta P_\eta - \bar\eta P_{\bar\eta})$, and
$H_0, S, \bar S$ and $F$ are defined in (\ref{40},\ref{41},\ref{42}). In
the component form of the action (\ref{61}) there are no kinetic terms
for $N, \psi, \bar\psi$ and $V$. This fact is reflected in the primary
constraints $P_N =0$, $P_\psi = 0$, $P_{\bar\psi} = 0$ and $P_V = 0$,
where $P_N, P_\psi, P_{\bar\psi}$ and $P_V$ are the canonical momenta
conjugate to $N, \psi, \bar\psi$ and $V$, respectively. Then, the total
Hamiltonian may be written as
\be
\tilde H = \tilde H_{c(n=2)} + u_NP_N + u_\psi P_\psi +
u_{\bar\psi} P_{\bar\psi} + u_V P_V.
\label{63}
\ee
Due to the conditions $\dot P_N = \dot P_\psi = \dot P_{\bar\psi} =
\dot P_V = 0$ we now have the first-class constraints
\begin{eqnarray}
\tilde H &=&  \frac{R^3}{\kappa^3}P_T + H_0 = 0, \qquad
{\cal F} = \frac{R^3}{\kappa^3} F_\eta + F = 0, \nonumber\\
Q_\eta &=& \frac{R^3}{\kappa^3} S_\eta + S = 0, \qquad
Q_{\bar\eta} = - \frac{R^3}{\kappa^3} S_{\bar\eta} + \bar S = 0.
\label{64}
\end{eqnarray}
They form a closed super-algebra with respect to the Dirac's brackets
\begin{eqnarray}
\lbrace Q_\eta, Q_{\bar\eta} \rbrace^\ast &=& - 2i \tilde H, \qquad
\lbrace \tilde H, Q_\eta \rbrace^\ast =
\lbrace \tilde H, Q_{\bar\eta} \rbrace^\ast = 0 \nonumber\\
\lbrace {\cal F}, Q_\eta \rbrace^\ast &=& iQ_\eta, \qquad
\lbrace {\cal F}, Q_{\bar\eta} \rbrace^\ast = -i Q_{\bar\eta}.
\label{65}
\end{eqnarray}
After quantization Dirac's brackets must be replaced by
anticommutators
\be
\lbrace \eta , P_\eta \rbrace = i\lbrace \eta, P_\eta \rbrace^\ast =
-\frac{\kappa^3}{R^3},\qquad
\lbrace \bar\eta, P_{\bar\eta}\rbrace =
i \lbrace \bar\eta, P_{\bar\eta} \rbrace^\ast = -\frac{\kappa^3}{R^3},
\label{66}
\ee
with the operator representation
\begin{eqnarray}
P_\eta &=& -\frac{\kappa^3}{R^3} \frac{\partial}{\partial\eta}, \qquad
P_{\bar\eta} = -\frac{\kappa^3}{R^3} \frac{\partial}{\partial\bar\eta} .
\label{67}
\end{eqnarray}
To obtain the quantum expression for $H_0, S, \bar S, F$ we
must solve the operator ordering ambiguity. Such ambiguities always take
place when the operator expression contains the product of non-commuting
operators $\lambda$ and $\bar\lambda$, $\chi$ and $\bar\chi$, $R$ and
$\pi_R = - i\frac{\partial}{\partial R}$, $\phi$ and $\pi_\phi = -i
\frac{\partial}{\partial\phi}$. Such procedure leads in our case to the
following expressions for the generators on the quantum level
\begin{eqnarray}
\tilde H &=&-i\frac{\partial}{\partial T} + H_0 (R,\pi_R,\phi , \pi_\phi
, \lambda , \bar\lambda ,\chi \bar\chi) , \nonumber \\
Q_\eta &=&
\Big(\frac{\partial}{\partial\eta}-i\bar\eta\frac{\partial}{\partial T}
\Big)+S(R,\pi_R ,\phi ,\pi_\phi ,\lambda ,\chi) , \label{68} \\
Q_{\bar\eta} &=& - \Big(-\frac{\partial}{\partial\bar\eta} +i\eta
\frac{\partial}{\partial T} \Big)+\bar S (R, \pi_R ,\phi ,\pi_\phi
,\bar\lambda ,\bar\chi ) ,\nonumber \\
{\cal F} &=& \Big(\eta \frac{\partial}{\partial\eta}-\bar\eta
\frac{\partial}{\partial\bar\eta} \Big) + F(\lambda ,\bar\lambda ,\chi
,\bar\chi ), \nonumber
\end{eqnarray}
where $S_\eta =\frac{\partial}{\partial\eta} - i\bar\eta
\frac{\partial}{\partial T}$ and $S_{\bar\eta} = -
\frac{\partial}{\partial\bar\eta} +i\eta
\frac{\partial}{\partial T}$ are the generators of the supertranslation,
$ P_T = -i \frac{\partial}{\partial T}$ is the ordinary time translation
on the superspace with coordinates $(t,\eta,\bar\eta)$
\be
\{ S_\eta , S_{\bar\eta} \} =2i \frac{\partial}{\partial T} ,
\label{69}
\ee
and $F_\eta = \eta \frac{\partial}{\partial\eta} - \bar\eta
\frac{\partial}{\partial\bar\eta}$ is the $ U(1)$ generator of the rotation
on the complex Grassmann coordinate $\eta (\bar\eta = \eta^\dagger)$.
The algebra of the quantum generators of the conserving charges $H_0 ,S,
\bar S, F$ is a closed super-algebra
\begin{eqnarray}
\lbrace S,\bar S \rbrace & = & 2H_0, \quad  \lbrack S, H_0\rbrack =
\lbrack \bar S,H_0 \rbrack = \lbrack F, H_0 \rbrack = 0,
\nonumber \\
S^2 =\bar S^2 &=&0,\qquad \lbrack F, S \rbrack = - S, \qquad
\lbrack F, \bar S \rbrack = \bar S.
\label{70}
\end{eqnarray}
We can see from Eqs. (\ref{65}) and (\ref{68}) that the operators
$\tilde H, Q_\eta , Q_{\bar\eta}$ and ${\cal F}$ obey the same super-algebra
(\ref{70})
\begin{eqnarray}
\lbrace Q_\eta ,Q_{\bar\eta}\rbrace &=& 2\tilde H,\quad
\lbrack Q_\eta , \tilde H \rbrack = \lbrack Q_{\bar\eta} ,\tilde H \rbrack =
\lbrack {\cal F}, \tilde H \rbrack = 0
\nonumber \\
Q^2_\eta &=& Q^2_{\bar\eta} = 0,\qquad [ {\cal F}, Q_\eta ]=- Q_\eta ,
\quad [{\cal F} , Q_{\bar\eta} ] = Q_{\bar\eta} .
\label{71}
\end{eqnarray}
In the quantum theory the first-class constraints (\ref{68}) become
conditions on the wave function $\Psi$, which has the superfield form
\begin{eqnarray}
\Psi (T,\eta,\bar\eta, R, \phi,\lambda,\bar\lambda,
\chi, \bar\chi ) &=& \psi (T, R, \phi,\lambda,\bar\lambda,\chi,\bar\chi )
\nonumber\\
&+& i\eta \xi (T,R,\phi ,\lambda ,\bar\lambda ,\chi ,\bar\chi )
+ i\bar\eta \zeta (T,R,\phi , \lambda ,\bar\lambda , \chi ,\bar\chi )
\nonumber\\
&+& \sigma (T,R,\phi , \lambda , \bar\lambda , \chi , \bar\chi )
\eta\bar\eta.
\label{72}
\end{eqnarray}

So, we have the supersymmetric quantum constraints
\be
\tilde H \Psi = 0, \qquad Q_\eta \Psi = 0, \qquad Q_{\bar\eta} \Psi = 0,
\qquad {\cal F} \Psi = 0.
\label{73}
\ee
As a consequence of the algebra (\ref{71}) the constraints
\be
Q_\eta \Psi = 0, \qquad Q_{\bar\eta} \Psi = 0,
\label{74}
\ee
lead to the equation
\be
\lbrace Q_\eta, Q_{\bar\eta} \rbrace \Psi = 2 \tilde H \Psi = 0,
\label{75}
\ee
which is a time-dependent Schr\"odinger equation for the
minisuperspace model.

The condition (\ref{75}) leads to the following form for the wave function
(\ref{72})
\be
\psi_\ast = \psi - \eta ( S \psi) - \bar\eta (\bar S \psi)+
\frac{1}{2} (\bar SS -S\bar S) \psi \eta\bar\eta ,
\label{76}
\ee
then $Q_\eta \psi_*$ has the following form
\begin{eqnarray}
Q_\eta \psi_* &=& \bar\eta \Big(-i\frac{d\psi}{dT} +\frac{1}{2} \{S,
\bar S\} \psi \Big)+ \nonumber \\
&+& \eta\bar\eta S\Big(-i\frac{d \psi}{dT}+\frac{1}{2} \{S,\bar S\}
\psi\Big) =0,
\label{77}
\end{eqnarray}
this is the standard Schr\"odinger equation and due to the relation
$H_0 =\frac{1}{2} \{ S,\bar S\}$ it may be written as
\be
i\frac{\partial \psi}{\partial T} = H_0 \psi,
\label{78}
\ee
where the wave function is
$\psi (T, R, \phi,\lambda ,\bar\lambda ,\chi ,\bar\chi )$. If we put in the
Schr\"odinger equation (\ref{78}) the condition of a stationary state
given by $\frac{\partial \psi}{\partial T} = 0$, we will have that
$H_0 \psi =0$ and due to the algebra (\ref{70}) we obtain $S \psi = \bar
S\psi = 0$ and the wave function $\psi_\ast$ becomes $\psi$.

The next step is to consider the additional term (\ref{30}) in the general
relativity theory and its consequences in the canonical formalism.

\vspace{.5cm}
\noindent {\bf Acknowledgments.} We are grateful to
E. Ivanov, S. Krivonos, J.L. Lucio, I. Lyanzuridi, L. Marsheva, O.
Obreg\'on, M.P. Ryan, J. Socorro and M. Tsulaia for their interest
in the work and useful comments. This research was supported in
part by CONACyT under the grant 28454E. Work of A.P. was supported
in part by INTAS grant 96-0538 and by the Russian Foundation of
Basic Research, grants 99-02-18417 and 99-02-04022(DFG). One of us
J.J.R. would like to thank CONACyT for support under Estancias
Posdoctorales en el Extranjero.


\end{document}